\documentclass[final,5p,times,twocolumn]{elsarticle}

\usepackage{graphicx}

\usepackage{amssymb}

\usepackage{amsmath}

\usepackage[dvips]{color}

\newcommand{\nn}{\nonumber}
\newcommand{\betaL}{\beta_{\mathrm{L}}}
\newcommand{\betaLC}{\beta_{\mathrm{L}}^{~\mathrm{c}}}
\newcommand{\betaRef}{\beta_{\mathrm{L}}^{~\mathrm{ref}}}

\journal{Physics Letters B}

\begin{document}

\begin{frontmatter}



\title{Chiral phase transition at finite temperature
and conformal dynamics \\
in large $N_f$ QCD}


\author[LNF]{Kohtaroh Miura}
\author[LNF,HU]{Maria Paola Lombardo}
\author[GU]{Elisabetta Pallante}

\address[LNF]{INFN Laboratori Nazionali di Frascati, 
	I-00044, Frascati (RM), Italy}
\address[HU] {Humboldt-Universit\"at zu Berlin, Institut f\"ur Physik, D-12489 Berlin, Germany} 
\address[GU]{Centre for Theoretical Physics, University of Groningen,
	9747 AG, Netherlands}

\begin{abstract}
We investigate the 
chiral phase transition
at finite temperature ($T$)
in colour SU$(N_c=3)$ Quantum Chromodynamics (QCD)
with a variable number  of fermions $N_f$ 
in the fundamental representation
by using  lattice QCD.
For $N_f=6$ we study the approach to asymptotic scaling 
by considering lattices 
with several temporal extensions $N_t$.
We then extract the dimensionless ratio
$T_c/\Lambda_{\mathrm{L}}$
($\Lambda_{\mathrm{L}}=$ Lattice Lambda-parameter)
for $N_f = 6$ and $N_f = 8$, the latter relying
on our earlier results. 
Further, we collect the (pseudo) critical couplings $\betaLC$
for the chiral phase transition at
$N_f=0$~(quenched), and $N_f = 4$ at a fixed $N_t = 6$. 
The results are consistent
with enhanced fermionic screening at larger $N_f$.
The ratio $T_c/\Lambda_{\mathrm{L}}$ depends very mildly 
on $N_f$ in the $N_f=0 - 4$ region, 
starts increasing at $N_f = 6$,
and becomes significantly larger at $N_f = 8$,
close to the edge of the conformal window.
We discuss interpretations of these results as well as their
possible interrelation with preconformal dynamics in the light
of a functional renormalization group analysis.
\end{abstract}

\begin{keyword}
Lattice Gauge Theory \sep Conformal Symmetry \sep Chiral Symmetry \sep Finite Temperature

\end{keyword}

\end{frontmatter}




{\em INTRODUCTION}\quad
Conformal invariance is anticipated to
emerge in asymptotically free  non-Abelian gauge theories
when the number of flavours exceeds a critical value $N_f={N_f}^c$.
The approach to conformality for $  N_f \lesssim {N_f}^c$, close
to the edge of the conformal window,  
is in principle associated with a walking behaviour of the running
coupling, which has been advocated as a basis for strongly interacting
mechanisms of electroweak symmetry breaking~\cite{Sannino:2009za}.

Recent lattice studies~\cite{DelDebbio:2010zz} 
focused on the computation of ${N_f}^c$ 
and the analysis of the conformal window itself, 
either with fundamental fermions
\cite{Appelquist:2011dp,Appelquist:Conformal,
Deuzeman:2008sc,Deuzeman:2009mh,Hasenfratz:2010fi, Hasenfratz:2009ea, 
Fodor:2011tu,Fodor:2009wk} or other representations 
\cite{Finland:MWT,Svetitsky:sextet,Kogut:2010cz,Fodor:2009ar}.
Among the many interesting results with fundamental fermions, we single
out the observation 
that QCD with three colours and eight flavours
is still in the hadronic phase
\cite{Appelquist:Conformal,Deuzeman:2008sc},
while $N_f = 12$ seems to be close to the critical number of flavours,
with some groups favouring conformality
\cite{Appelquist:2011dp,Appelquist:Conformal,
Deuzeman:2009mh,Hasenfratz:2010fi}, 
and others chiral symmetry breaking~\cite{Fodor:2011tu}.

In comparison,  much less effort has been devoted to the analysis of the
phenomenologically relevant subcritical region
\cite{Appelquist:2009ka,Appelquist:2010xv,Catterall:MWT,Lucini:MWT}.
Here, we would like to learn where and 
how QCD at the edge of the conformal
window displays walking,
and its associated manifestations like separation of scales
and approximate scale invariance.
This Letter is one step in this direction.

Building on the above mentioned results,
we decided to concentrate ourselves
on $N_f \le 8$, so to be safely in the hadronic region,
but not too far from the edge of the conformal window. 
A recent study~\cite{Appelquist:2009ka} noted an
enhancement of the zero temperature ratio
$\langle \bar{\psi} \psi \rangle / F^3 $,
where $F$ is the pseudoscalar decay constant.
This suggests that $N_f = 6$
might indeed be the onset of new strong dynamics.

In this letter we study the thermal transition of
QCD with $N_f$ = {0,4,6},
and combine our findings with those of
our early work for $N_f=8$ \cite{Deuzeman:2008sc}.
We confirm the expected enhanced screening when the number of flavours
increases, and discuss the interrelation of our results 
with a possible  emergence of a new, preconformal dynamics.

As a general remark, we note that 
using the thermal transition
as a tool for investigating preconformal dynamics 
was largely inspired by a renormalization group analysis
\cite{BraunGies}, as we will
review below.  Further reasons of interest for finite temperature studies 
of large $N_f$ include a connection between the quark-gluon plasma
phase and the cold conformal region, which might lend support to 
analyses of quark-gluon plasma based on the AdS/CFT correspondence.

{\em BRIEF OVERVIEW OF ANALYTIC RESULTS}\quad
A second zero of the two-loop
beta-function of a non-Abelian gauge theory
implies, at least perturbatively,
the appearance of an infrared
fixed point (IRFP)
and the restoration of conformal symmetry
\cite{Caswell:1974gg,Banks:1981nn}.
In colour SU($3$) gauge
theory with $N_f$ massless fermions in the fundamental
representation, the second zero appears
for $N_f\gtrsim 8.05$, before 
the loss of asymptotic freedom (LAF) at
$N_f^{\mathrm{LAF}}=16.5$.

One expects that
conformality should emerge
when the renormalized coupling at the would be IRFP
is not strong enough to break chiral symmetry.
This condition provides 
the lower bound $N_f^c$ of a so called conformal window in $N_f$.  
Analytic studies based on the Schwinger-Dyson equation with
rainbow resummations
\cite{Appelquist:1996dq,Miransky:1997,Appelquist:1999hr}
or the functional renormalization group method
~\cite{BraunGies}  
suggest $N_f^c\sim 12$.
An all-order perturbative beta-function
\cite{Ryttov:2007cx}
inspired by the NSVZ
beta-function of SQCD \cite{Novikov:1983uc} has been conjectured,
leading to a bound $N_f^c > 8.25$;
$N_f^{c}$ has also been estimated
for different fermion representations \cite{Dietrich:2006cm}.
In addition to the lower bound of the conformal window,
walking dynamics in the preconformal region
is another interesting subject
in relation to proposed scenarios of walking technicolor.
Instanton studies at large $N_f$ \cite{Velkovsky:1997fe}
claimed a qualitative change of behaviour at $N_f=6$.

All the above phenomena happen well
into the strong coupling regime,
rendering a perturbative prediction unreliable
and a non-perturbative analysis
mandatory. The genuinely non-perturbative 
lattice formulation of gauge theories
is thus a natural candidate for this study.

Recently, the functional renormalization group (FRG)  method
has been applied to finite
$T$ QCD with varying number of flavours,
and the critical temperature
for the chiral phase transition
was obtained as a function of $N_f$~\cite{BraunGies}.
In this $T-N_f$ phase diagram,
the onset of the conformal window has been estimated
by locating the vanishing critical temperature.

Most interestingly, the critical exponents associated with 
the behaviour of the beta function at the IRFP
manifest themselves also in the shape of the thermal critical
line in the vicinity of the critical number of flavours $N_f^c$. 
In more detail,
the line is almost linear with $N_f$ 
for small $N_f$, and displays a singular behaviour when
approaching $N_f^c$.
As emphasised by the authors in Ref.~\cite{BraunGies},
the result clearly elucidates the universality of the critical
behaviour at zero and non-zero temperature in the vicinity of $N_f^c$.
It thus seems a promising direction
to extend the knowledge of finite $T$ lattice QCD
to the larger $N_f$ region, 
by using the FRG results as analytic guidance.

In this work 
we investigate the thermal chiral phase transition
for $N_f=0,4,6,8$ colour SU$(N_c=3)$ QCD
by using lattice QCD Monte Carlo simulations
with staggered fermions. $N_f=6$ is expected to be in the important regime
as suggested by the results
in Refs.~\cite{Appelquist:2011dp,Velkovsky:1997fe}.
This work includes the first study of 
$N_f=6$ staggered fermions at finite $T$, and it provides
an important ingredient to a broader project that studies
the emergence of the conformal window
in the $T-N_f$ phase diagram.
In addition to $N_f=6$, we compute the (pseudo) critical coupling
for $N_f=0$ (quenched) and $N_f=4$ at $N_t = 6$,
and use the results from Ref.~\cite{Deuzeman:2008sc} for $N_f=8$.

In short, this work explores a largely uncharted territory: 
the chiral transition of strong interactions at high temperature,
and large number of light flavours.
Our goal is to observe, and understand
possible qualitative differences 
with the very well known behaviour of QCD thermodynamics.
To this end, we have for the first time collected and analysed
results for different number of flavours,
and the same lattice action,
so to be able to meaningfully compare them.  
We ask the question:
are we still finding just small
differences among theories with different number of flavours,
or are we going to observe some significant trend?
As we will see, our exploratory,
and in many respects qualitative analysis,
will indicate that $N_f=6$, and even more $N_f=8$,
are serious candidates for a different chiral dynamics.
Obviously, our observations call for detailed quantitative
studies which are already underway.


{\em SETUP}\quad
Simulations have been performed by utilising
the publicly available MILC code~\cite{MILC}.
The setup explained below is the same
as the one used for $N_f=8$ in Ref.~\cite{Deuzeman:2008sc}.
We use an improved version of the staggered action, the
Asqtad action, with a one-loop Symanzik
\cite{Bernard:2006nj,LuscherWeisz}
and tadpole~\cite{LM1985} improved gauge action,
\begin{align}
S = -\frac{N_f}{4}\mathrm{Tr}\log M[am,U,u_0]
+ \sum_{i=p,r,pg}\beta_i(g^2_{\mathrm{L}})
\mathrm{Re}\bigl[1-U_{C_i}\bigr]\ ,\label{eq:action}
\end{align}
where $g_{\mathrm{L}}$ is the lattice bare coupling,
and $\beta_i$ are defined as
\begin{align}
&
\bigl(
\beta_p,\beta_r,\beta_{pg}
\bigr)
=
\biggl(
\frac{10}{g_{\mathrm{L}}^2},
-\frac{\beta_p(1-0.4805\alpha_s)}{20u_0^2},
-\frac{\beta_p}{u_0^2}0.03325\alpha_s
\biggr)\ \label{eq:beta} \\
&
\alpha_s=-4\log\frac{u_0}{3.0684}\ ,\quad
u_0=\langle U_{C_p}\rangle^{1/4}\ .
\end{align}
The plaquette coupling
$\beta_p=10/g_{\mathrm{L}}^2\equiv \beta_{\mathrm{L}}$
is a simulation input.
The $M[am,U,u_0]$ in Eq.~(\ref{eq:action})  denotes the matrix
for a single flavour Asqtad fermion with bare lattice mass $am$,
and $U_{C_i}$ represents the trace of
the ordered product of link variables along $C_i$,
for the $1\times 1$ plaquettes ($i=p$),
the $1\times 2$ and $2\times 1$ rectangles ($i=r$),
and the $1\times 1\times 1$ parallelograms ($i=pg$),
respectively -  all divided by the number of colours.
The tadpole factor $u_0$ is determined
by performing zero temperature simulations
on the $12^4$ lattice, and used as an input
for finite temperature simulations.

To generate configurations with mass degenerate
dynamical flavours,
we have used the rational hybrid Monte Carlo algorithm
(RHMC)~\cite{Clark:2006wq}, which allows to 
simulate an arbitrary number of
flavours through varying the number of pseudo-fermions.
Simulations for $N_f=6$ have been performed by using two
pseudo-fermions, and subsets of trajectories for
the chiral condensates and Polyakov loop
have been compared with those obtained
by using three pseudo-fermions with the same
Monte Carlo time step $d\tau$
and total time length $\tau$ of a single trajectory.
We have observed very good
agreement between the two cases
for both evolution and thermalization.
We have monitored the Metropolis acceptance and reject ratio,
and adjusted $\tau=0.2 - 0.24$ and $d\tau=0.008 - 0.018$
to realize the best performance.
For each parameter set,
we have collected a number of trajectories
ranging from a one thousand to five thousand
- the latter closer to the critical region.

The focus of this letter
is the analysis of the chiral transition.
The fundamental observable is then the order parameter for
chiral symmetry, the chiral condensate:
\begin{equation}
a^3\langle\bar{\psi}\psi\rangle =
\frac{N_f}{4N_s^3N_t}
\Big\langle\mathrm{Tr\bigl[M^{-1}\bigr]}\Big\rangle
\ ,\label{eq:PBP}
\end{equation}
where $N_s~(N_t)$ represents the number of lattice sites
in the spatial (temporal) direction and
$U_{4,t\mathbf{x}}$ is the temporal link variable.
We have also measured connected and disconnected chiral
susceptibilities,
\begin{align}
a^2\chi_{\mathrm{conn}} &= 
-\frac{N_f}{4 N_s^3 N_t}
\langle \mathrm{Tr} \left[( MM )^{-1}\right ] \rangle
\ ,\nonumber\\
a^2\chi_{\mathrm{disc}} &=
\frac{N_f^2}{16 N_s^3 N_t}
\left [  \langle \mathrm{Tr} \left[M^{-1}\right] ^2\rangle
- \langle \mathrm{Tr} \left[M^{-1}\right] \rangle^2\right ]\ ,
\end{align}
and we have considered the logarithmic derivative,
\begin{equation}
R_\pi \equiv \chi_\sigma/\chi_\pi\ ,
\label{eq:R_pi} 
\end{equation}
where,
\begin{equation}
\chi_\sigma \equiv\chi
= \frac {\partial \langle \bar \psi \psi\rangle}{\partial m}
= \chi_\mathrm{conn} + \chi_\mathrm{disc}\ ,
\end{equation}
and
\begin{equation}
\chi_\pi = \frac {\langle \bar \psi \psi\rangle}{m}\, .
\end{equation}
As discussed in previous work
\cite{Deuzeman:2008sc,Kocic:1992is},
$R_\pi$ is a probe of chiral symmetry
which is particularly useful for numerical investigations.
To further characterize the critical region,
we also measured the Polyakov loop,
\begin{equation}
L =
\frac{1}{N_cN_s^3}\sum_{\mathbf{x}}
\mathrm{Re}
\bigg\langle
\mathrm{tr}_c\prod_{t=1}^{N_t}U_{4,t\mathbf{x}}
\bigg\rangle
\ ,\label{eq:PLOOP}
\end{equation}
and $\mathrm{tr}_c$ denotes the trace in colour space.

The temperature $T$ is related to 
the inverse of the lattice temporal extension,
\begin{align}
&T\equiv \frac{1}{a(\beta_{\mathrm{L}})\cdot N_t}\ .\label{eq:T}
\end{align}
We measure $\langle\bar{\psi}\psi\rangle$,
the chiral susceptibilities and $L$ at various temperatures.
The output of this measurement
is the (pseudo) critical coupling $\betaLC$ for
the chiral phase transition for a given value of $N_t$.
We underscore that 
all the measurements of the pseudo-critical couplings 
which will be used in our discussion
are only based on fermionic observables.

\begin{figure}[ht]
\includegraphics[width=8.0cm]{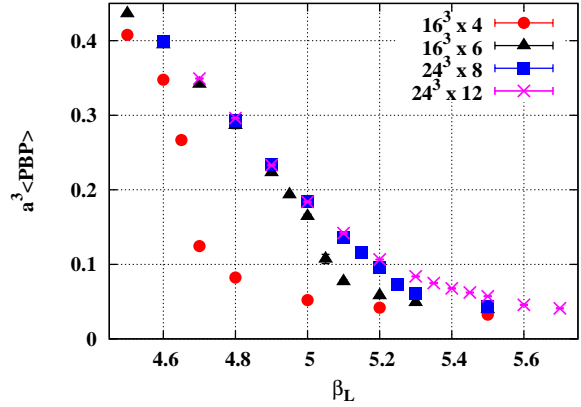}
\caption{The chiral condensate
$a^3\langle\bar{\psi}\psi\rangle$
for $N_f=6$ and $am=0.02$ in lattice units,
as a function of $\beta_{\mathrm{L}}$,
for $N_t=4,~6,~8$, and $12$.
Error-bars are smaller than symbols.}
\label{Fig:PBP_beta}
\end{figure}

\begin{figure}[ht]
\includegraphics[width=8.0cm]{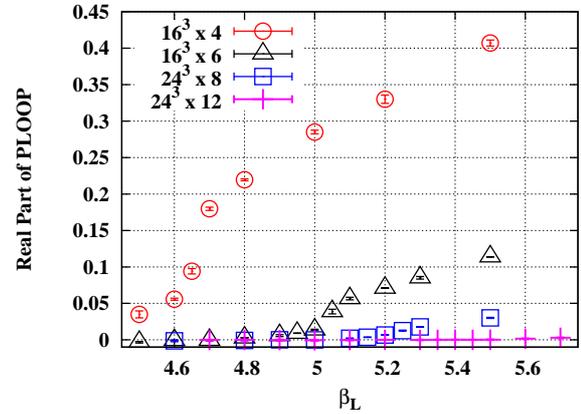}
\caption{The Polyakov loop $L$
for $N_f=6$ and $am=0.02$ in lattice units,
as a function of $\beta_{\mathrm{L}}$,
for $N_t=4,~6,~8$, and $12$.
Error-bars are smaller than symbols.}
\label{Fig:PLOOP_beta}
\end{figure}

{\em RESULTS}\quad
All results have been obtained for 
a fermion bare lattice mass $am=0.02$.
In Figs.~\ref{Fig:PBP_beta} and \ref{Fig:PLOOP_beta},
the expectation values of
the chiral condensate $a^3\langle\bar{\psi}\psi\rangle$,
and the Polyakov loop $L$ are displayed
as a function of $\beta_{\mathrm{L}}$ for several $N_t$,
respectively.It is found that different $N_t$
give a different behaviour of
$a^3\langle\bar{\psi}\psi\rangle$ and $L$.
In particular, this indicates that their rapid crossover
with increasing $\beta_{\mathrm{L}}$ is not to be attributed to
a bulk  transition. The asymptotic scaling analysis
below will confirm that it corresponds instead to 
a thermal chiral phase transition (or crossover)
in the continuum limit.
\footnote{A note on the mass dependence is in order.
According to the Pisarski-Wilczek scenario,
the most likely possibility for $N_f \ge 3$ 
is a first order chiral transition in the chiral limit.
Standard arguments indicate that first
order phase transitions are robust against explicit breaking:
when introducing a bare quark mass, then,
we expect a first order phase transition which
will eventually end in a genuine singularity
at some critical point ($T_c, m_c)$.
By further increasing the bare mass,
the transition will turn into a crossover.
In the unexpected situation of a second order transition
in the chiral limit, any nonzero quark mass will immediately produce
a crossover. Even in the case of a first order transition, though,
a finite lattice will turn it  into a crossover.
All in all, as in any lattice study,
we are never dealing directly with a genuine criticality.
Rather, we are locating a pseudo-critical point,
and only by considering several masses and several volumes,
we can assess with confidence
the nature of the phase transition
in the infinite volume and in the chiral limit.
Discriminating among these different behaviours is however
beyond the scope of this study.}

For $N_t=4$,
it is possible to extract $\betaLC=4.65(25)$
from the peak position of the first derivative
$-\, a^3 d\langle\bar{\psi}\psi\rangle/d\beta_{\mathrm{L}}$.
We note that at this level of accuracy we cannot disentangle
the peak position for the derivatives of the chiral condensate
and the Polyakov loop $dL/d\beta_{\mathrm{L}}$.

For $N_t=6$,
we find a small jump between $\beta_{\mathrm{L}}=5.0$ and $5.05$,
where the Polyakov loop also shows a significant
enhancement.

\begin{figure}[ht]
\includegraphics[width=8.0cm]{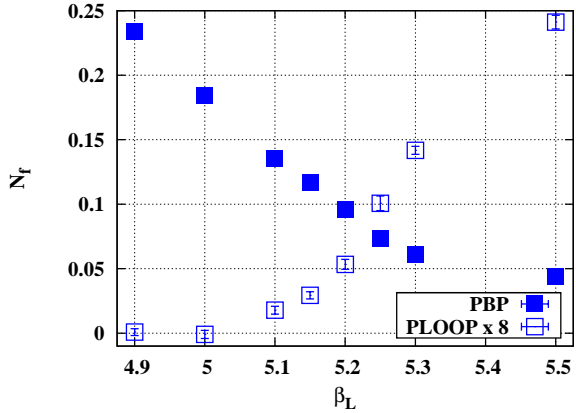}
\caption{Zoom-in of the chiral condensate
$a^3\langle\bar{\psi}\psi\rangle$ and the Polyakov loop
$L$ shown in Figs.~\protect\ref{Fig:PBP_beta}
and \protect\ref{Fig:PLOOP_beta} in the critical
region at $N_t=8$, with spatial volume $24^3$.}
\label{Fig:PBP_PLOOP_beta}
\end{figure}

\begin{figure}[ht]
\includegraphics[width=8.0cm]{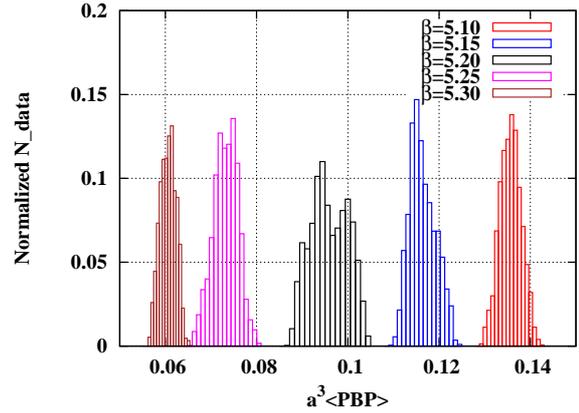}
\caption{Distribution of the chiral condensate
$a^3\langle\bar{\psi}\psi\rangle$
for $N_f=6$, $am=0.02$ and spatial volume $24^3$,
in the vicinity of the chiral phase transition at $N_t=8$.}
\label{Fig:PBP_HG_Nt8}
\end{figure}

For $N_t=8$, the Polyakov loop $L$
shows a clear signal
as indicated in Fig.~\ref{Fig:PBP_PLOOP_beta}.
In particular, we observe a drastic increase of
the Polyakov loop $L$
around $5.2< \beta_{\mathrm{L}} < 5.3$.
The histogram of the chiral condensate
around the rapid increase of the Polyakov loop
is shown in Fig.~\ref{Fig:PBP_HG_Nt8}.
At $\beta_{\mathrm{L}}=5.2$,
the histogram exhibits a broadening
that suggests the increase of fluctuations around the
pseudo-critical point.  

For $N_t=12$ the results are particularly smooth. 
Note that in this
case the aspect ratio is only two,
and larger volumes would be required
to reach a comparable clarity in the signal, although
the onset for the Polyakov loop
at $\beta = 5.50(5)$ is still appreciable.

\begin{figure}[ht]
\includegraphics[width=8.0cm]{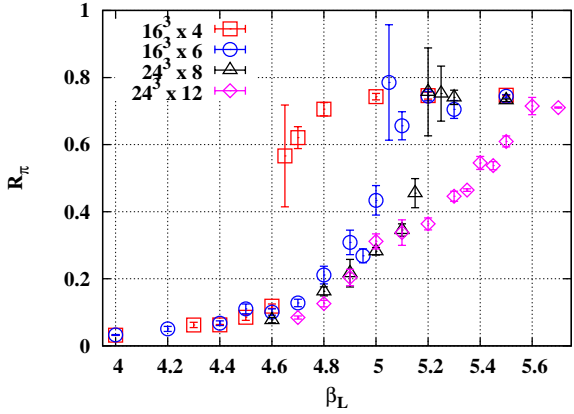}
\caption{The ratio of scalar and pseudo-scalar contributions
to the susceptibility, defined in Eq.~(\protect\ref{eq:R_pi})
as a function of $\betaL$.}
\label{Fig:Rpi}
\end{figure}

The results for $R_\pi$ are collected in Fig.\ref{Fig:Rpi}.
$R_\pi$ has a clear jump for
$\beta_c = (4.65(5),5.05(5),5.20(5))$ for $N_t = (4,6,8)$. 
For $N_t =12$, the behaviour is again rather smooth.
We have then searched for an inflection point
in $R_\pi$ by fitting to
a hyperbolic tangent in several intervals,
varying the extrema of the integrations.
The results are reasonably stable in the errors. 
We will quote as a central value the average of the fit
results within different intervals, and for the error,
we will use the conservative estimate of half the
difference between the largest and smallest result,
obtaining $\beta_c = 5.46(14)$.
All values of
the (pseudo) critical lattice coupling $\betaLC$
are summarized in Table~\ref{Tab:bc}.

These results can be analyzed and interpreted in terms of
the two-loop asymptotic scaling law.
Let us consider the two-loop lattice beta function,
\begin{align}
&\beta({g})
=-(b_0 {g}^3 + b_1 {g}^5)\ ,\label{eq:beta_func}\\
&b_0
=
\frac{1}{(4\pi)^2}
\Biggl(
\frac{11C_2[G]}{3}-\frac{4T[F]N_f}{3}
\Biggr)\ ,\label{eq:b0}\\
&b_1
=
\frac{1}{(4\pi)^4}
\Biggl(
\frac{34(C_2[G])^2}{3}
-\biggl(\frac{20C_2[G]}{3}+4C_2[F]\biggr)T[F]N_f
\Biggr)\ ,\label{eq:b1}
\end{align}
for fundamental fermions in $SU(N_c)$
$(C_2[G],\,C_2[F],\,T[F])=(N_c,\,(N_c^2-1)/(2N_c),\, 1/2)$.
From  Eq.~(\ref{eq:beta_func})
we obtain the well known two-loop asymptotic scaling law,
\begin{align}
\Lambda_{\mathrm{L}}~a(\beta_{\mathrm{L}})
&=
\biggl(\frac{2N_cb_0}{\betaL}\biggr)^{-b_1/(2b_0^2)}
\exp\biggl[
\frac{-\betaL}{4N_cb_0}
\biggr]
\ .\label{eq:2Loop_AS_Lat}
\end{align}
Here $\Lambda_{\mathrm{L}}$ is
the so called lattice Lambda-parameter,
and $\beta_L = 2 N_c/ g^2$.

The above relation is valid in the massless limit.
In the following,
we will use it to analyze results obtained
at finite mass. This assumes that the shift of the 
(pseudo) critical coupling induced by a non-zero mass
is smaller than other errors.
This assumption should ultimately
be tested by performing simulations with different masses and
extrapolating to the chiral limit.

The equation~(\ref{eq:T}) can be written as
\begin{align}
\frac{1}{N_t}=\frac{T_c}{\Lambda_{\mathrm{L}}}
\times \Bigl(\Lambda_{\mathrm{L}}~a(\betaLC)\Bigr)
\ .\label{eq:TcLambda}
\end{align}
The left-hand side is a given number,
and we have obtained
the corresponding $\betaLC$ by lattice simulations.
Hence, the quantity $T_c/\Lambda_{\mathrm{L}}$
on the right-hand side of Eq.~(\ref{eq:TcLambda})
can be extracted,
and must be unique
as long as the asymptotic scaling law
Eq.~(\ref{eq:2Loop_AS_Lat})
is verified for a given $\betaLC$.

If we use the lattice bare coupling to carry out
the above program, we notice appreciable scaling violations.
Indeed, Eq.~(\ref{eq:2Loop_AS_Lat}) holds true
up to non-universal scaling-violating terms. 
One possibility is then to follow earlier work \cite{Cheng:2007jq}
and to parametrize scaling violations as
\begin{align}
%
{\Lambda}_{\mathrm{L}}~a(\beta_{\mathrm{L}})
\equiv
\bigl[
\Lambda_{\mathrm{L}}a(\beta_{\mathrm{L}})
\bigr]_{\mathrm{2loops}}
\Bigl(
1 + h \bigl[
\Lambda_{\mathrm{L}}a(\beta_{\mathrm{L}})
\bigr]_{\mathrm{2loops}}^2
\Bigr)\ .\label{eq:2Loop_AS_Lat_BF}
\end{align}
where $\bigl[
\Lambda_{\mathrm{L}}a(\beta_{\mathrm{L}})
\bigr]_{\mathrm{2loops}}$
is defined by Eq.~(\ref{eq:2Loop_AS_Lat}).

Alternatively, we can trade the bare lattice coupling
$g_{\mathrm{L}}$ for the boosted coupling
introduced in our previous work~\cite{Deuzeman:2008sc}
$g = \sqrt{2N_c/10}\cdot g_{\mathrm{L}}$.
We will show below that this prescription leads to a
rather accurate two-loop scaling,
equivalent to the enhancement of the scaling behaviour
obtained by considering Eq.~(\ref{eq:2Loop_AS_Lat}). 
Our prescription is similar in spirit
to the Parisi-Lepage-Mackenzie~\cite{LM1985}
boosted coupling.
Our boosted coupling, however,
cannot be derived in (tadpole improved)
perturbation theory.
In our future work,
we plan to perform zero temperature
measurements which will allow an independent
estimate of the lattice spacing,
and a complete discussions of scaling,
and asymptotic scaling.
Hopefully, this will
shed light on the reasons why our simple approach works
so well. At this stage, admittedly,
it remains a heuristic, ad-hoc 
prescription which effectively incorporates the scaling
violations in the range we have explored. 
This warning issued, we will continue the discussion
by using our boosted coupling.

\begin{figure}[ht]
\includegraphics[width=8.0cm]{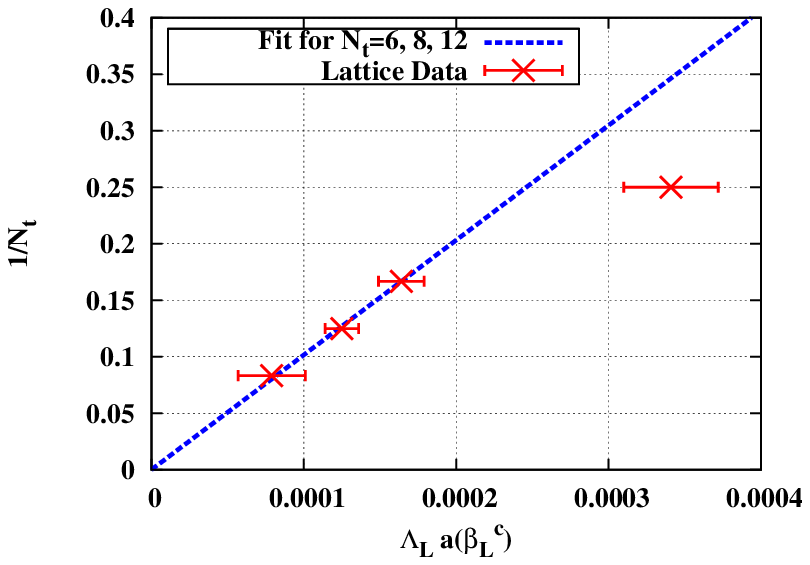}
\caption{The thermal scaling behaviour of the 
(pseudo) critical lattice coupling $\betaLC$.
Data points for $\Lambda_{\mathrm{L}}~a(\betaLC)$
at a given $1/N_t$ are obtained by using $\betaLC$
from Table~\protect\ref{Tab:bc} as input for extracting
$\Lambda_{\mathrm{L}}~a(\betaLC)$
in the two-loop expression
Eq.~(\protect\ref{eq:2Loop_AS_Lat}).
The dashed line is a linear fit
with zero intercept to the data with $N_t>4$.}
\label{Fig:NtScale}
\end{figure}

In Fig.~\ref{Fig:NtScale},
we show the $N_t^{-1}-\Lambda_{\mathrm{L}}a(\betaLC)$ plot.
The slope of the line connecting the origin
and the data points corresponds to  $T_c/\Lambda_{\mathrm{L}}$.
The $N_t=6,~8$, and $12$ points have
a common slope to a very good approximation,
while the $N_t=4$ result falls on a smaller slope.

The latter is interpreted as
a scaling violation effect due to the use of a too small $N_t$.
The existence of a common $T_c/\Lambda_\mathrm{L}$ for $N_t\geq 6$
indicates that the data are consistent with the two-loop asymptotic scaling
Eq.~(\ref{eq:2Loop_AS_Lat}), 
confirms the thermal nature of the transition and that $N_f=6$ is 
outside the conformal window, as expected from a previous $N_f=8$ study
\cite{Deuzeman:2008sc}.
A linear fit provides 
$T_c/\Lambda_{\mathrm{L}} = 1.02 (12) \times  10^3 $,
which can be interpreted as the value in the continuum limit
for $N_f=6$ QCD.
Note that this compares very well
with the result obtained by using 
$\betaLC$ obtained by $N_t=8$ simulations.
In the following,
we can then use it as a representative result
for six flavours.

In order to have a more complete overview,
we have performed simulations for the theory
with $N_f=0$ (quenched) and $N_f=4$, only at 
$N_t = 6$. These theories are of course very well investigated,
however we have not found
in the literature results for the same action as ours.
Table~\ref{Tab:bc} shows a summary of our results for 
the (pseudo) critical coupling  $\betaLC$
of the chiral phase transition at finite temperature 
for $N_f=0,~4,~6$, and $8$ - 
the latter from  Ref.~\cite{Deuzeman:2008sc}.


\begin{figure}[ht]
\includegraphics[width=8.0cm]{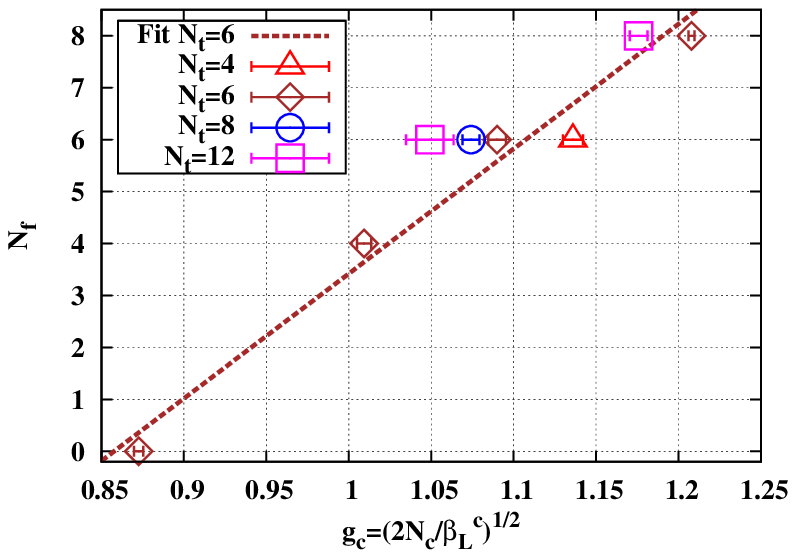}
\caption{(Pseudo) critical values of the lattice coupling
$g_c=\sqrt{2N_c/\betaLC}$ for theories with $N_f=0,~4,~6,~8$ 
and for several values of $N_t$ in the Miransky-Yamawaki phase diagram.
The dashed (brown) line is a linear fit to the $N_t=6$ results.}
\label{Fig:MY}
\end{figure}

{\em DISCUSSION}\quad
In Fig.~\ref{Fig:MY},
we display the (pseudo) critical values of the lattice coupling
$g_c=\sqrt{2N_c/\betaLC}$ from Table \ref{Tab:bc}
in the Miransky-Yamawaki phase diagram.

Consider the $N_t = 6$ results:
it is expected that an
increasing number of flavours favors
chiral symmetry restoration.
Indeed, we find that, on a fixed lattice,
the (pseudo) critical coupling increases with $N_f$
in agreement with early studies and naive reasoning. 
The precise dependence of 
the (pseudo) critical coupling on $N_f$ at fixed $N_t$
is not known.
It is, however,
amusing to note that the results seem to be
smoothly connected by an almost straight line:
the brown line in the plot is a linear fit to the data.
Comparing the trend for $N_f = 6$ to the one for $N_f = 8$,
for varying $N_t$
one can infer
a decreasing in magnitude (and small) step scaling function,
hence a walking behaviour.
Further study is needed at larger $N_f$,
and by using the same action used for $N_f=0-8$,
to confirm or disprove it.

\begin{table*}[ht]
\caption{
Summary of the (pseudo) critical lattice couplings $\betaLC$
for the theories with $N_f=0,~4,~6,~8$, $am=0.02$
and varying $N_t=4,~6,~8,~12$.
All results are obtained using the same lattice action.
}\label{Tab:bc}
\begin{center}
\begin{tabular}{c|c|c|c|c}
\hline
$N_f\backslash N_t$ &
$4$&
$6$&
$8$&
$12$\\
\hline
$0$ &
-&
$7.88\pm 0.05$&
-&
-\\
$4$ &
-&
$5.89\pm 0.03$&
-& 
\\
$6$ &
$4.65\pm 0.05$&
$5.05\pm 0.05$&
$5.2\pm 0.05$&
$5.45\pm 0.15$\\
$8$ &
-&
$4.1125\pm 0.0125$&
-&
$4.34\pm 0.04$\\
\hline
\end{tabular}
\end{center}
\end{table*}

Next, we study 
the $N_f$ dependence of the ratio
$T_c/\Lambda_{\mathrm{L}}$ and related quantities.
We recall that
the simulations for $N_f = 4$ and $N_f = 0$
have been performed by using only $N_t = 6$.
Hence, in these two cases, 
the results will hold true barring strong scaling
violations at $N_t = 6$. We note 
that in a previous lattice study
with improved staggered fermions~\cite{Gupta:2000hr},
asymptotic scaling was indeed observed
using a boosted coupling 
for $N_t \ge 6$ for $0 \le N_f \le 4$.

Ideally, we would like to convert our results
to $T_c/\Lambda_{\bar{\mathrm{MS}}}$.
Unfortunately, to our knowledge,
the conversion from
$\Lambda_{\mathrm{L}}$ to
$\Lambda_{\bar{\mathrm{MS}}}$
for a generic number of flavours
is only available for Wilson fermions~\cite{Kawai:1980ja}.

Here we consider a simplified procedure,
aiming at capturing
at least the basic features induced by setting a UV scale.
For this purpose,
we introduce a reference coupling $\betaRef$ and 
an associated reference energy scale $\Lambda_{\mathrm{ref}}$.
Then Eq.~(\ref{eq:2Loop_AS_Lat}) is generalized as
\begin{align}
&\Lambda_{\mathrm{ref}}(\betaRef)~a(\betaL)\nn\\
&=
\Biggl(
\frac{b_1}{b_0^2}~
\frac{\betaL+2N_cb_1/b_0}
{\betaRef+2N_cb_1/b_0}
\Biggr)^{b_1/(2b_0^2)}\exp
\Biggl[-\frac{\betaL-\betaRef}{4N_cb_0}
\Biggr]\ .
\label{eq:int_as}
\end{align}
At leading order of perturbation theory $b_1\to 0$,
$\Lambda_{\mathrm{L}}$ and $\Lambda_{\mathrm{ref}}$
are related via
\begin{align}
\frac{\Lambda_{\mathrm{ref}}}{\Lambda_{\mathrm{L}}}
=\exp\Biggl[\frac{\betaRef}{4N_cb_0}\Biggr]
\ .\label{eq:L_ratio}
\end{align}
This equation would be analogous of the ratio
$\Lambda_{\mathrm{L}}/\Lambda_{\mathrm{MS}}$
derived in \cite{Kawai:1980ja}
for Wilson fermions up to a further linear
dependence on $N_f$ in the numerator of the exponent.
In a nutshell, the difference originates
from the fact that
we are fixing a bare reference coupling
$\betaRef$, which will be specified later.
Notice that
by construction $\Lambda_{\mathrm{ref}}$
reproduces the lattice Lambda-parameter
$\Lambda_{\mathrm{L}}$ in the limit
\begin{align}
\Lambda_{\mathrm{ref}}(
\betaRef\to 0)
=\Lambda_{\mathrm{L}}
\Bigl(
1+\mathcal{O}\bigl(1/\betaLC\bigr)
\Bigr)\ .
\end{align}
In summary, when trading $\Lambda_{\mathrm{L}}$ for
$\Lambda_{\mathrm{ref}}$,
we are moving towards a more UV scale.

\begin{figure}[ht]
\includegraphics[width=8.0cm]{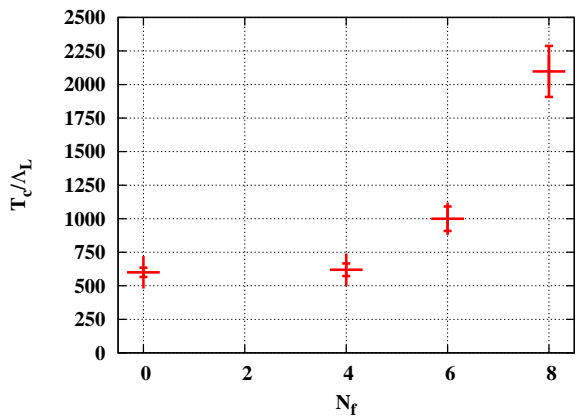}
\caption{The ratio $T_c/\Lambda_{\mathrm{L}}$,
for $N_f=0,~4,~6$ and 8 and lattice bare mass $am=0.02$.}
\label{Fig:TcL_LP}
\end{figure}

Let us consider first
$T_c/\Lambda_{\mathrm{L}}$. 
The values of $T_c/\Lambda_{\mathrm{L}}$
are summarized in Table \ref{Tab:TcL},
and plotted in Fig.~\ref{Fig:TcL_LP}.
The ratio does not show a significant $N_f$ dependence
in the region $0\leq N_f\leq 4$, it  
starts increasing at $N_f = 6$,
and undergoes a rapid rise around $N_f=8$.
The chiral phase transition would happen
when $T$ becomes comparable to a typical energy scale
$M_{\chi} = C \Lambda_{\mathrm{L}}$.
The nearly constant nature of $T_c/\Lambda_{\mathrm{L}}$
in the region $N_f\le 4$ indicates that
the role of such  energy scale is not significantly
changed by the variation of $N_f$ (see  \cite{Braun:2009si}
for a detailed discussion of this point.)
In turn, the increase of
$T_c/\Lambda_{\mathrm{L}}$ in the region $N_f\ge 6$
might well imply that the chiral dynamics
becomes different from the one for $N_f\leq 4$.
Indeed, a recent lattice study~\cite{Appelquist:2009ka}
indicates that $N_f = 6$ is close to
the threshold for preconformal dynamics.

\begin{figure}[ht]
\includegraphics[width=8.0cm]{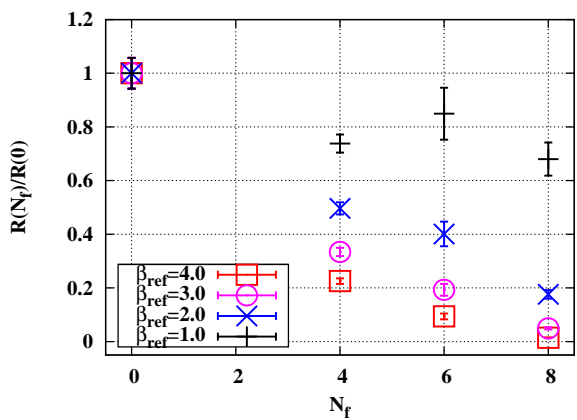}
\caption{The $N_f$ dependence of
$R(N_f)/R(0)$ for several finite fixed $\betaRef$.
Here, $R(N_f)\equiv (T_c/\Lambda_{\mathrm{ref}})(N_f)$.
The limit $\betaRef\to 0$
reproduces the results shown
in Fig.~\protect\ref{Fig:TcL_LP}
up to a renormalization factor and
and up to a corrections 
$\mathcal{O}\bigl(1/\betaLC\bigr)$.}
\label{Fig:TcL}
\end{figure}

We now consider $T_c/\Lambda_{\mathrm{ref}}$. 
The $N_f$ dependence of the ratio 
$R(N_f)\equiv (T_c/\Lambda_{\mathrm{ref}})(N_f)$ is shown
for several $\betaRef$
in Fig.~\ref{Fig:TcL}, where the vertical axis
is normalized by $R(0)=(T_c/\Lambda_{\mathrm{ref}})(N_f=0)$
for each $\betaRef$.
$T_c/\Lambda_{\mathrm{ref}}$ is now
a decreasing function of $N_f$
for a larger $\betaRef$, {\em i.e.}
for a more UV reference scale $\Lambda_{\mathrm{ref}}$.
This result is consistent with the FRG study~\cite{BraunGies},
where the decreasing $T_c(N_f)$
{has been obtained by using the $\tau$-lepton mass
$m_{\tau}$} as a common UV reference scale
with a common coupling $\alpha_s(m_{\tau})$.

The $\Lambda_{\mathrm{ref}}$ scale
associated with a $\betaRef\gg\beta_*$
where $\beta_*$ is evaluated at
the infra-red fixed point
should provide a UV scale
well-separated from the IR dynamics.
If we assume the lower bound of the conformal
window to be $N_f^c\simeq 12$,
the two-loop beta-function
leads to $\beta_*=-2N_cb_1/b_0\simeq 0.63$.
Indeed Fig.~\ref{Fig:TcL} shows that
the decreasing nature of
$(T_c/\Lambda_{\mathrm{ref}})(N_f)$
is still weak at $\betaRef = 1.0$.
In the limit $\betaRef\to 0$,
$T_c/\Lambda_{\mathrm{ref}}$ reproduces Fig.~\ref{Fig:TcL_LP},
and the resultant increasing feature
should be attributed to
the vanishing of $\Lambda_{\mathrm{L}}$
due to infra-red dynamics.
We also notice that $\betaRef$
must always be smaller than $\beta$ at the UV cutoff,
$\beta_{\mathrm{UV}}=\betaLC(N_f)$.
As shown in Table \ref{Tab:bc},
the lowest value of the (pseudo) critical coupling is given by
$\betaLC(N_f=8,N_t=6)=4.1125\pm 0.0125$,
hence we constrain our analyses
to $\betaRef\leq 4.0$.
In summary, Figs.~\ref{Fig:TcL_LP} and~\ref{Fig:TcL} together
show the effects of shifting the reference scales
from the IR to the UV.

With the use of a UV reference scale,
we should observe the predicted critical behavior
\cite{BraunGies}
\begin{align}
T_c(N_f) = K |N_f - N_f^c|^{-1/ \theta}\ .
\end{align}
By choosing the critical exponent
$\theta$ in the range predicted by FRG:
$1.1 < 1/|\theta| < 2.5$,
our data are consistent with the values  
$N_f^c = 9(1)$ for $\betaRef = 4.0$ and $N_f^c = 11(2)$ for
$\betaRef = 2$. We plan to extend and refine this
analysis in the future, and here we only notice a reasonable
qualitative behaviour.

\begin{table}[ht]
\caption{
$T_c/\Lambda_{\mathrm{L}}$ for several $N_f$. 
Results are obtained by using the same lattice
action.
For $N_f=6$, we have used the $N_t=8$ result
as a representative value.
The values for $N_f=8$ are extracted
from Ref.~\cite{Deuzeman:2008sc}.
}\label{Tab:TcL}
\begin{center}
\begin{tabular}{c|c}
\hline
$N_f$ &
$T_c/\Lambda_{\mathrm{L}}$ \\
\hline
$0$ &
$600\pm 34$ \\
$4$ &
$620\pm 28$ \\
$6$ &
$1000\pm 92$ \\
$8$ &
$2098 \pm 191$\\
\hline
\end{tabular}
\end{center}
\end{table}

{\em SUMMARY}\quad
We have investigated the chiral phase transition
and its asymptotic scaling
for $N_f=6$ colour SU$(3)$ QCD
by using lattice QCD Monte Carlo simulations
with improved staggered fermions.
This study provides an important ingredient
to a broader project that studies
the emergence of a conformal window
in the $T-N_f$ phase diagram.
We have determined the (pseudo) critical lattice coupling
$\betaLC$ for several lattice temporal extensions
$N_t$. We have extracted the dimensionless ratio
$T_c/\Lambda_{\mathrm{L}}$
($\Lambda_{\mathrm{L}}=$Lattice Lambda-parameter)
for the theory with $N_f=6$
using two-loop asymptotic scaling.
The analogous result for $N_f=8$ has been extracted
from Ref.~\cite{Deuzeman:2008sc}.
$T_c/\Lambda_{\mathrm{L}}$ 
for $N_f=0$ and $N_f=4$ has been measured at fixed $N_t=6$,
barring asymptotic scaling violations.
Then we have discussed the $N_f$ dependence of the ratios
$T_c/\Lambda_{\mathrm{L}}$ and $T_c/\Lambda_{\mathrm{ref}}$,
where $\Lambda_{\mathrm{ref}}$ is a UV reference energy scale, 
related to $\Lambda_{\mathrm{L}}$ as in Eq.~(\ref{eq:L_ratio}).

We have observed that $T_c/\Lambda_{\mathrm{L}}$
shows an increase in the region 
$N_f=6-8$, while it is approximately
constant in the region $N_f\leq 4$. 
We have discussed this qualitative change
for $N_f\geq 6$ and a possible relation with a preconformal phase.
We repeat that all results have been
obtained by working at one value of the quark mass
and this is a potential weakness of our calculations.

The ratio $T_c/\Lambda_{\mathrm{ref}}$
is a decreasing function of $N_f$.
This behaviour is consistent with
the result obtained
in the functional renormalization group analysis~\cite{BraunGies},
where a common UV reference scale
was used to study
the chiral phase boundary
in the $T-N_f$ phase diagram.

Next steps of the current project involve a scale setting
at zero temperature by measuring a common UV observable.
It would also be desirable to have the relation between
$\Lambda_{\mathrm{L}}$ and $\Lambda_{\bar{\mathrm{MS}}}$
for our action.

This, together with a more extended set of flavour
numbers, will allow a quantitative analysis of the critical
behaviour. 
We expect the resultant $T_c-N_f$ phase diagram
to play an essential role in the study of
the conformal window.

{\em ACKNOWLEDGMENTS}\quad
We thank Holger Gies and Jens Braun
for fruitful discussions and most useful suggestions. We have
enjoyed discussing these topics with Koichi Yamawaki, Masafumi Kurachi,
Hiroshi Ohki,  
Michael M\"uller-Preussker, Marc Wagner, Biagio Lucini,
Volodya Miransky, Albert Deuzeman and Tiago Nunes da Silva.
Kohtaroh Miura thanks Michael M\"uller-Preussker
and the theory group
in the Humboldt University for their hospitality.
Kohtaroh Miura is partially supported by
EU I3HP2-WP22.
This work was in part based on the MILC Collaboration's public
lattice gauge theory code. See 
http://www.physics.indiana.edu/\~ {}sg/milc.html
for details.
The numerical calculations were carried out on the
IBM-SP6 at CINECA,
Italian-Grid-Infrastructures in Italy,
and the Hitachi SR-16000 at YITP, Kyoto University in Japan.





\end{document}